\documentclass[useAMS,usenatbib,referee]{mn2e}{{ 
\usepackage{graphicx} 
\usepackage{natbib} 
\include{epsf} 
\include{rotate} 
\title[Enhanced formation of HD and D$_2$ on dust]{Enhanced  
production of HD and 
D$_2$ molecules on small dust grains  
in diffuse clouds 
} 
 
\author[A. Lipshtat, O. Biham and E. Herbst]{Azi Lipshtat$^{1}$, Ofer  
Biham$^{1}$ and Eric Herbst$^{2}$ \\   
$^{1}$Racah Institute of Physics, The Hebrew University, Jerusalem  
91904, Israel \\ 
$^{2}$Departments of Physics, Astronomy, and Chemistry, 
 The Ohio State University, Columbus, OH  
43210, USA 
} 
 
\begin{document} 
\maketitle 
 
\begin{abstract} 
Motivated by recent observations of deuterated molecules  
in the interstellar medium, we examine the   
production of HD and D$_2$ molecules on dust grain surfaces. 
A mechanism for the enhanced production of these deuterated  
molecules is studied. This mechanism applies on grain surfaces  
on which D atoms stick more strongly than H atoms,  
under conditions of low flux  
and within a suitable range of temperatures. 
It is shown that under these conditions  
the production rates of HD and D$_2$ are greatly enhanced  
(vs. the H$_2$ production rate)  
compared with the expected rates based on the adsorption of  
gas-phase atomic abundances of D and H.   
The enhancement  
in the formation rate of HD is comparable with the enhancement due to  
gas-phase ion-molecule reactions in diffuse clouds. 
\end{abstract} 
 
\begin{keywords} 
ISM: molecules - molecular processes 
\end{keywords} 
 
\section{Introduction} 
        \label{intro.} 
 
The formation of molecular hydrogen  
on the surfaces of dust grains 
in the interstellar medium 
is a process of fundamental importance  
because the formation rate in the gas 
phase cannot account for the observed  
abundance 
\citep{Gould1963}.  
Once produced on grain surfaces and ejected or desorbed into the gas,  
hydrogen molecules are a necessary component in the  
initiation of gas-phase chemical reaction networks that  
give rise to the chemical complexity observed in  
interstellar clouds. 
The formation of molecular hydrogen has  
been studied both theoretically 
\citep{Hollenbach1970, Hollenbach1971a, Hollenbach1971b, Williams1968,  
Smoluchowski1981, Smoluchowski1983, Aronowitz1985, Duley1986,  
Pirronello1988, 
Sandford1993,Takahashi1999}  
and experimentally 
\citep{Brackmann1961, Schutte1976,  
Pirronello1997a, Pirronello1997b, Pirronello1999, Manico2001}.   
Although formation of H$_{2}$ is clearly the most studied surface  
reaction pertaining to interstellar clouds, there are many other  
reactions that are thought to take place, for the most part involving  
H atoms and other species.  Reactions are typically  
considered to occur via the Langmuir-Hinshelwood mechanism, in which  
weakly-bound surface species diffuse towards one another.  Some  
authors have considered different mechanisms, where chemisorption  
plays a role \citep{Cazaux02}. 
  
The computational modeling of diffusive 
chemical reactions on dust grains in the interstellar medium 
is typically done using rate 
equation models 
\citep{Pickles1977, d'Hendecourt1985,  Brown1990, Brown1990b,   
Hasegawa1992,  Hasegawa1993a, Hasegawa1993b,  Caselli1993,   
Willacy1993,  Shalabiea1994}.  
As long as a sizable number of reactive species exists on a  
grain surface,  
the rate equation results are valid.  
However, in the limit of small  
average surface populations ($\le  
1$) of reactive species, 
rate equations are not always suitable,  because they 
take into account only average densities and ignore the fluctuations  
as well as the discrete nature of the populations of the atomic and  
molecular species 
\citep{Tielens1995,Tielens1982,Charnley1997,Caselli1998,Shalabiea1998,Stantcheva2001,  
Stantcheva2002}.  
Depending  
on the parameters chosen, this limit can pertain mainly to  
small grains or can even to ``classical'' grains, namely 
grains of 0.1$\mu$ radius.   
For example, as the number of H atoms, $N_{\rm H}$, on a grain  
fluctuates  
in the range of 0, 1 or 2, 
the H$_2$ formation rate cannot be obtained from the average  
number alone. 
This can be easily understood, since  recombination  requires at  
least two 
H atoms simultaneously on the surface. 
 
Recently, a master equation approach was proposed that  
is suitable for the simulation of diffusive chemical reactions on  
interstellar 
grains in the accretion limit 
\citep{Biham2001,Green2001,Biham2002}. 
The master equation 
takes into account both the discrete nature of the 
reactive species as well as the fluctuations in their populations.  
\citet{Green2001} considered a simple system in which H and O  
atoms land on classical grains to produce H$_2$, OH,  and O$_2$.  
 Biham and co-workers focused on  
the formation of H$_2$.   
With a slow diffusion rate for H based on  
the analysis by \citet{Katz1999} of experimental work by  
Pirronello, Vidali, \& co-workers 
\citep{Pirronello1997a,Pirronello1997b}, 
they found the master equation approach to  
be needed mainly for small grains.    
For the formation of H$_2$,  
the dynamical variables are the 
probabilities $P_{\rm H}(N_{\rm H})$  
that there are  
$N_{\rm H}$  
atoms 
on the grain 
at time $t$.  
The time derivatives   
$\dot{P}_{\rm H}(N_{\rm H})$, $N_{\rm H}=0, 1, 2, \dots,$ 
are expressed  
in terms of the adsorption, reaction and desorption terms. 
The master equation provides the time evolution of  
$P_{\rm H}(N_{\rm H})$,  
$N_{\rm H}=0, 1, 2, \dots$,  
from which 
the recombination rate can be calculated. 
The master equation has also been used for 
the study of 
reaction networks involving multiple species 
\citep{Stantcheva2002,Stantcheva2003}. 
 
In addition to the master equation method, a Monte Carlo 
method \citep{Tielens1982} was 
proposed 
\citep{Charnley2001}, 
and applied to surface reaction networks involving multiple 
species 
\citep{Caselli2002,Stantcheva2003}. The weakness of the Monte Carlo  
approach to surface chemistry is that it cannot be used easily for  
models in which gas-phase reactions take place simultaneously.   
Finally, a semi-empirical approach, known as the modified rate method, was  
proposed by \citet{Caselli1998}.  Although artificial, the modified rate method is easy to  
employ and has been used with mixed success \citep{ Stantcheva2001,  
Caselli2002}. 
 
In addition to H, which is the most 
abundant atom in the universe, there is  
a density of D atoms, which is smaller by a  
factor of about $10^{-5}$ compared with the density of H atoms in 
regions where the gas is primarily atomic. 
If deuterated hydrogen molecules, namely HD and D$_2$, are 
formed on grain surfaces in the same way as 
H$_2$, these molecules are expected to be rare due to the low  
abundance of D atoms. 
Of course, some differences between the formation rates of  
HD and D$_2$ and that of H$_2$ are expected because of the 
possibly different diffusion and desorption rates of 
H and D atoms on the surface. It was realized early in the study  
of interstellar chemistry, however, that the formation of HD in  
diffuse interstellar clouds occurs more efficiently in the gas phase  
via the sequence of reactions 
 
\begin{eqnarray} 
{\rm H}^{+}  &+&  {\rm D}  \longrightarrow {\rm D}^{+}  +  {\rm H}  
\nonumber \\ 
{\rm D}^{+}  &+&  {\rm H}_{2} \longrightarrow {\rm HD}  +  {\rm H}^{+}, 
\end{eqnarray} 
 
\noindent 
which can lead to enhancements of $\approx 10^{2}$ over surface  
formation \citep{Watson1973,Dalgarno1973} and explain the observed HD/H$_{2}$  
abundance ratio in diffuse clouds.  By the stage in which dense  
interstellar clouds have been formed, HD has become the major  
repository of deuterium, and is the starting point for a rich  
chemistry of deuterium fractionation that occurs in the gas phase  
via ion-molecule reactions  \citep{Watson1973, Millar1989, Charnley1997,Roberts2000a,  
Roberts2000b,Rodgers2001, Roberts2002}. 
 Since  
gas-phase fractionation produces a large D/H atomic abundance ratio  
in dense clouds, it then becomes possible that surface processes  
can produce multiply deuterated species in high abundance following  
the adsorption of D atoms \citep{Charnley1997,Rodgers2001, Caselli2002,Stantcheva2003}. 
 
Although it has been thought for some time that the formation of  
HD in diffuse clouds occurs primarily in the gas, the surface  
processes for the formation of HD and D$_{2}$ have not been looked at  
carefully. In this paper we propose a mechanism taking place on the 
surfaces of dust grains that, under certain  
conditions, can provide a significantly enhanced production of  
deuterated 
molecules. 
We demonstrate the mechanism for the simple case that involves 
only H and D atoms in diffuse clouds, 
and examine the formation rates of HD and D$_2$ 
vs. that of H$_2$. 
The mechanism is based on the assumption that D atoms stick to 
the surface more strongly than H atoms so that their desorption 
rate is lower. 
An isotope effect of this type has been observed in various 
experimental situations 
\citep{Koehler1988,Hoogers1995}. 
  
Due to the very low cosmic abundance ratio of D vs. H  
($10^{-5}$) most of the atoms that hit the grain  
surfaces in diffuse clouds are H atoms.  
However, the residence time of D atoms on grains 
is expected to be longer than that of H atoms, due to their 
lower desorption rate.  
Thus, the D/H atomic abundance ratio on the grains is enhanced  
compared with the gas-phase ratio. 
As a result, newly adsorbed H (or D) atoms are more likely to find D atoms 
already residing on the grains, 
giving rise to an enhanced production of HD molecules. 
This enhancement turns out to be even larger on very small grains 
on which only a handful of H and D atoms typically reside.  
This effect is expected to be significant  
in diffuse clouds, where the density of the gas phase is low and 
the grain size distribution is broad and dominated by very small grains  
\citep{Li2001}. 
 
The paper is organized as follows.  
The interaction between hydrogen atoms and dust grains in  
interstellar clouds in described in Sec. 2. 
The formation of molecular hydrogen on large grains  
is described in Sec. 3 using the rate equation model.  
The formation of molecules on small grains is presented 
in Sec. 4 using the master equation approach, followed by 
an asymptotic analysis for very small grains in Sec. 5. 
The simulations and results are presented in Sec. 6. 
The results are discussed in Sec. 7 
and summarised in Sec. 8. 
In Appendix A we consider the relevance of the isotope  
effect studied here to related experiments. 
 
\section{The interaction between hydrogen atoms and dust grains} 
 
Consider a diffuse interstellar cloud dominated by a density  
$n_{\rm H}$ (cm$^{-3}$) of H atoms and also including a density  
$n_{\rm D} = f n_{\rm H}$ of D atoms, where $f$ is the D$/$H  
abundance ratio (the cosmic ratio is approximately $f=10^{-5}$).  The  
typical velocity of the hydrogen isotope I (H or D) in the gas phase  
is denoted by $v_{\rm I}$ (cm s$^{-1}$).  It is given by 
 
\begin{equation} 
v_{\rm I} = \sqrt{ {8 \over \pi} {k_B T_{\rm gas} \over m_{\rm I}} }, 
\label{eq:velocity} 
\end{equation} 
 
\noindent where the mass $m_{\rm I}$ is $m_{\rm H} = 1.67 \times  
10^{-24}$ (grams) for H atoms and $m_{\rm D} = 3.34 \times 10^{-24}$  
(grams) for D atoms and the temperature of the gas is denoted by  
$T_{\rm gas}$. 
The cloud includes a density  
$n_{\rm gr}$ (cm$^{-3}$) of dust grains.  To evaluate the flux of  
atoms onto grain surfaces we will assume, for simplicity, that the  
grains are spherical with a radius $r$ (cm).  The cross section of a  
grain is $\sigma = \pi r^2$ and its surface area is $4 \sigma$.  The  
number of adsorption sites on a grain is denoted by $S$, and their  
density $s$ (sites cm$^{-2}$) on the surface is given by $s=S/(4  
\sigma)$.  The fluxes $F_{\rm H}$ (atoms s$^{-1}$) and $F_{\rm D}$ of  
H and D atoms onto the surface of a single grain are given by $F_{\rm  
I} = n_{\rm I} v_{\rm I} \sigma$.  The atoms stick to the surface  
and hop as random walkers between adjacent sites until they either  
desorb as atoms or recombine into molecules.  The desorption rates of  
H and D atoms on the surface are given by 
 
\begin{equation} 
W_{\rm I} =  \nu \cdot \exp [- E_{1}({\rm I}) / k_{B} T],   
\label{eq:P1} 
\end{equation} 
 
\noindent 
where $\nu$ is the attempt rate  
(typically assumed to be $10^{12}$ s$^{-1}$),  
$E_{1}({\rm I})$  
is the energy needed for desorption  
of an atom of isotope I and $T$ is the surface temperature. 
The hopping rates of the atoms from site to site are 
 
\begin{equation}  
a_{\rm I} =  \nu \cdot \exp [- E_0({\rm I}) / k_{B} T],  
\label{eq:Alpha} 
\end{equation} 
 
\noindent where $E_0({\rm I})$ is the activation energy barrier for  
diffusion of the isotope I. Here we assume that diffusion occurs only  
by thermal hopping, in agreement with recent experimental results  
\citep{Katz1999}.   
In the faster diffusion rates often adopted by  
astrochemists, tunneling is allowed to occur \citep{Hasegawa1992} and  
results in very different rates for H and D. The  
rate $A_{\rm I} = a_{\rm I}/S$ is approximately the inverse of the  
time $t_s({\rm I})$ required for an atom of isotope I to visit nearly  
all the adsorption sites on the grain surface, because in two  
dimensions the number of distinct sites visited by a random walker is  
linearly proportional to the number of steps, up to a logarithmic  
correction \citep{Montroll1965}. 
 
In this paper, we take the density of adsorption sites on the surface  
as $s = 5 \times 10^{13}$ (sites cm$^{-2}$) \citep{Biham2001}.  The  
density of H atoms is $n_{\rm H} = 50$ (cm$^{-3}$) and the density  
of D atoms is $n_{\rm D} = f n_{\rm H}$, where $f = 10^{-5}$ is  
the cosmic abundance ratio, which according to Eq.~(\ref{eq:velocity})  
gives rise to to the ratio $F_{\rm D}/F_{\rm H} = 10^{-5}/\sqrt{2}$.  
The gas temperature is taken as $T_{\rm gas}=90$ K,  
thus $v_H=1.37 \times 10^5$ and $v_D=9.7 \times 10^4$ (cm s$^{-1}$).   
The surface temperature is $T=18$ K, 
although variations from this 
value are considered. 
These values are within the range  
of typical parameters for diffuse interstellar clouds. 
 
The parameters used for the interaction between the adsorbed H and D  
atoms and the grain are crucial to the determination of the  
isotope effect.  Since D is heavier than H, it sits lower than H 
in any potential surface when zero-point energy is included.  
The smaller zero-point energy of D leads to larger values for the  
desorption energy and the barrier against diffusion,  
although the differences between H and D are uncertain.   
In the analysis of their experimental  
results, for example,  \citet{Katz1999} assumed no  
difference between H and D atoms at all.   
In a previous paper, \citet{Caselli2002}  
argued for a difference of 2 meV between both the desorption energies  
and barriers against diffusion.  Here, the adopted energy barriers for  
diffusion are $E_0(H)=35$ meV for H atoms and $E_0(D)=37$ meV for D  
atoms, while the desorption energies are $E_1(H)=50$ meV for H  
atoms and $E_1(D)=60$ meV for D atoms.  The values for H  
atoms are intermediate between those obtained experimentally on  
olivine and amorphous carbon surfaces \citep{Katz1999}.  The  
difference between the desorption energies is assumed to be 10 meV; an  
argument in favour of this larger value is that the barrier against  
diffusion balances the zero-point energy of a potential well with  
that of a saddle point while desorption does not possess a saddle point.  
In any case, the difference  
in desorption energies between H and D is crucial to the isotope  
effect discussed here. 
 
\section{Molecular hydrogen formation on large grains} 
 
After being adsorbed onto the surface, the H and D atoms 
hop between adsorption sites until they either  
desorb or meet one another and form molecules. 
The number of atoms of isotope I on the surface of a grain is  
denoted by 
$N_{\rm I}$. 
The rate equation analysis deals with the 
expectation values  
$\langle N_{\rm I} \rangle$, I $=$ H, D 
obtained by averaging over a large ensemble of identical 
grains  
under given physical conditions. 
The rate equations  
for the expectation values of the H and D 
populations 
take the form 
 
\begin{eqnarray} 
{ {d{ \langle N_{\rm H}  \rangle }} \over {dt}}  &=&   
F_{\rm H}   
- W_{\rm H}  \langle N_{\rm H}  \rangle  
- 2 A_{\rm H}  {\langle N_{\rm H}  \rangle}^{2}  
-  (A_{\rm H}+A_{\rm D})   
{\langle N_{\rm H} \rangle \langle N_{\rm D}  \rangle} \nonumber \\  
{ {d{ \langle N_{\rm D}  \rangle }} \over {dt}}  &=&   
F_{\rm D}   
- W_{\rm D}  \langle N_{\rm D}  \rangle  
- 2 A_{\rm D}  {\langle N_{\rm D}  \rangle}^{2} 
-(A_{\rm H} + A_{\rm D})\langle N_{\rm H}\rangle \langle  N_{\rm D}  
\rangle.  
\label{eq:NHgrain} 
\end{eqnarray} 
 
\noindent 
These equations provide the time derivatives of the  
sizes of the populations of reactive species. 
Each equation includes a flux term, a desorption term 
and two reaction terms. The first reaction term 
accounts for reactions between atoms of the same specie, 
while the second reaction term accounts for the reaction 
that produces HD molecules. 
The rate of the latter reaction is proportional to the rate 
at which H and D atoms encounter each other, which 
is proportional to the sum of the hopping rates 
of the two species.  
Therefore, the pre-factor of this 
term is the sum of  
$A_{\rm H}$ 
and 
$A_{\rm D}$. 
By integration of the rate equations  
for a given initial condition 
one obtains the time dependence of the population sizes.  
The rate equations provide a good description of the  
surface reactions with our chosen parameters as long as the grain  
is not too small and the flux is not too low. 
The rate equations are 
strictly valid when both the D and H populations per grain are  
greater than unity. However, we have found that, despite a very small D  
population in even our largest grains, as long as the H  
population per grain is not too much smaller than unity,  
the rate equations provide a reasonable approximation to the master 
equation results.  
 
We will now solve the rate equations for the case of steady state  
conditions, 
in which 
$d \langle N_{\rm I} \rangle / d t = 0$  
for I $=$ H, D. 
Due to the small ratio between the fluxes of $D$ and $H$ atoms,  
the effect of the D atoms on the density  
$\langle N_{\rm H}  \rangle$ 
can be neglected. 
Thus, neglecting the HD formation term in the first equation in 
(\ref{eq:NHgrain}) 
we obtain 
 
\begin{equation} 
\langle N_{\rm H}  \rangle =  
{1 \over 4} \left( {W_{\rm H} \over A_{\rm H}} \right) 
\left[  
-1 + \sqrt{1 + 8 {(F_{\rm H}/W_{\rm H}) \over (W_{\rm H}/A_{\rm H})}  
}    
\right]. 
\label{eq:steadyNH} 
\end{equation} 
 
\noindent 
Inserting this solution into the second equation in 
(\ref{eq:NHgrain}) 
we obtain 
 
\begin{equation} 
\langle N_{\rm D}  \rangle = {1 \over 4} \left( {W_{\rm eff} \over  
A_{\rm D}} \right) 
\left[  
-1 + \sqrt{1 + 8 {(F_{\rm D}/W_{\rm eff}) \over (W_{\rm eff}/A_{\rm  
D})} }    
\right], 
\label{eq:steadyND} 
\end{equation} 
 
\noindent 
where  
 
\begin{equation} 
W_{\rm eff} = W_{\rm D} +(A_{\rm H}+A_{\rm D}) \langle N_{\rm H}   
\rangle  
\label{eq:Weff} 
\end{equation} 
 
\noindent 
is the effective desorption coefficient for the D atoms.  
The ratio  
$\langle N_{\rm D} \rangle / \langle N_{\rm H} \rangle$ 
between the population sizes of D and H on the 
grain surface is obtained from Eqs. 
(\ref{eq:steadyNH})  
and   
(\ref{eq:steadyND}). 
This ratio turns out to be higher than 
the gas phase ratio  
$D/H$ 
due to the lower desorption rate of D atoms. 
The production rate of  
HD molecules per grain is 
 
\begin{equation} 
R_{\rm HD} = (A_{\rm H}+A_{\rm D})    
\langle N_{\rm H}  \rangle \langle N_{\rm D}  \rangle  
\label{eq:HD} 
\end{equation} 
 
\noindent 
and of  
$H_2$ molecules is 
 
\begin{equation} 
R_{\rm H_2}=A_{\rm H}  \langle N_{\rm H}  \rangle^2.  
\label{eq:HH} 
\end{equation} 
 
\noindent 
The ratio between these production  
rates is thus 
 
\begin{equation} 
{R_{\rm HD} \over R_{\rm H_2}} =  
\left( 1+{A_{\rm D}\over A_{\rm H}}\right)  
{ \langle N_{\rm D} \rangle \over \langle N_{\rm H} \rangle }. 
\label{eq:ProdRatio} 
\end{equation} 
 
\noindent 
Due to the assumption that 
$E_0(H) < E_0(D)$, 
namely that 
the diffusion of H atoms is faster 
than of D atoms, we obtain that the additional enhancement factor 
multiplying 
$\langle N_{\rm D} \rangle / \langle N_{\rm H} \rangle$  
is in the range 
$1 < 1 + A_{\rm D}/A_{\rm H} < 2$. 
The production rate of D$_2$ molecules is 
 
\begin{equation} 
R_{\rm D_2}=A_{\rm D}  \langle N_{\rm D}  \rangle^2,  
\label{eq:DD} 
\end{equation} 
 
\noindent 
and the ratio between the D$_2$ and H$_2$ production rates 
is

\begin{equation} 
{R_{\rm D_2} \over R_{\rm H_2}} =  
{A_{\rm D} \over A_{\rm H}}  
{ \langle N_{\rm D} \rangle^2 \over \langle N_{\rm H} \rangle^2 }. 
\label{eq:ProdRatioD2H2} 
\end{equation} 
 
\section{Molecular hydrogen formation on small grains}

The master equation that describes the hydrogen-deuterium system  
consists of a two-dimensional matrix of equations. 
These equations describe the 
time derivatives of 
the probabilities  
$P(N_{\rm H},N_{\rm D})$ 
of having  
$N_{\rm H}$ 
hydrogen atoms and 
$N_{\rm D}$ 
deuterium atoms 
simultaneously on the surface. 
The master equation takes the form 
 
\begin{eqnarray} 
\label{eq:NmicroHD} 
\dot P(N_{\rm H},N_{\rm D}) &=&  
F_{\rm H} [P(N_{\rm H}-1,N_{\rm D})  
-  P(N_{\rm H},N_{\rm D})] \nonumber \\  
&+&F_{\rm D} [P(N_{\rm H},N_{\rm D}-1)  
- P (N_{\rm H},N_{\rm D}) ] 
\nonumber\\ 
&+& W_{\rm H} [ (N_{\rm H}+1) P (N_{\rm H}+1,N_{\rm D})  
-N_{\rm H} P (N_{\rm H},N_{\rm D})]  \nonumber\\ 
&+& W_{\rm D} [ (N_{\rm D}+1) P (N_{\rm H},N_{\rm D}+1)  
- N_{\rm D} P (N_{\rm H},N_{\rm D})]  \\  
&+&  A_{\rm H} [ (N_{\rm H}+2)(N_{\rm H}+1) P (N_{\rm H}+2,N_{\rm D})  
- N_{\rm H}(N_{\rm H}-1)P (N_{\rm H},N_{\rm O}) ] \nonumber \\ 
&+&  A_{\rm D} [(N_{\rm D}+2)(N_{\rm D}+1) P (N_{\rm H},N_{\rm D}+2)  
- N_{\rm D}(N_{\rm D}-1) P (N_{\rm H},N_{\rm D}) ]\nonumber\\ 
&+&  (A_{\rm H}+A_{\rm D})  
[(N_{\rm H}+1)(N_{\rm D}+1) P (N_{\rm H}+1,N_{\rm D}+1) -  
N_{\rm H} N_{\rm D} P (N_{\rm H},N_{\rm D}) ], \nonumber 
\end{eqnarray} 
 
\noindent 
where 
$N_{\rm H},N_{\rm D}=0,1,2,\dots$. 
Each of these equations includes three sets of terms. 
The first set describes the effect of the incoming fluxes $F_H$ and $F_D$  
on the probabilities. 
The probability  
$P(N_{\rm H},N_{\rm D})$  
increases when an H atom sticks to the surface of a grain that already 
has $N_{\rm H}-1$ adsorbed H atoms and $N_{\rm D}$ adsorbed D atoms. 
There is a similar term for the adsorption of a D atom.   
The probability 
$P(N_{\rm H},N_{\rm D})$  
decreases when a new atom is adsorbed on a grain with 
$N_{\rm H}$ H atoms  
and 
$N_{\rm D}$ D atoms  
that are already adsorbed. 
The second set of terms describes the effect of desorption.  
An H [D] atom that is 
desorbed from a grain on which there are  
$N_{\rm H}$ adsorbed H atoms   
and 
$N_{\rm D}$ adsorbed D atoms   
decreases the 
probability  
$P(N_{\rm H},N_{\rm D})$ 
and increases the probability 
$P(N_{\rm H}-1,N_{\rm D})$ 
[$P(N_{\rm H},N_{\rm D}-1)$]. 
The third set includes three terms describing  
the reactions that lead to 
the formation of 
H$_2$, D$_2$ and HD, respectively. 
The rate of each of these reactions is proportional to the 
sum of the sweeping rates 
of the atoms involved 
($A_{\rm H}$  
or  
$A_{\rm D}$)  
times the number of pairs of atoms that may participate 
in the reaction. 
In the equations in which  
$N_{\rm H}=0,1$ 
or  
$N_{\rm D}=0,1$, 
some of the terms vanish. 
These terms can be easily identified since their 
initial or final states include a negative number 
of H or D atoms on the grain.

By suitable summation over the probabilities,  
the master equation provides the moments of the population 
size distribution, given by

\begin{equation} 
\langle N_{\rm H}^k N_{\rm D}^{\ell}  \rangle =  
\sum_{N_{\rm H}=0}^{\infty}  
\sum_{N_{\rm D}=0}^{\infty}  
N_{\rm H}^k  
N_{\rm D}^{\ell}  
P(N_{\rm H},N_{\rm D}), 
\label{eq:momentkl} 
\end{equation} 
  
\noindent 
where $k,\ell = 0,1,2,\dots$ 
[note that in case that $k=0$ ($\ell=0$) the corresponding 
power,  
$N_{\rm H}^k$  
($N_{\rm D}^{\ell}$),  
is removed from the equation]. 
The order of each moment is specified by  
the sum of these powers, namely by 
$m = k + \ell$. 
Clearly, there are two first-order moments, 
$\langle N_{\rm H} \rangle$ 
and  
$\langle N_{\rm D} \rangle$, 
three second-order moments, 
$\langle N_{\rm H}^2 \rangle$, 
$\langle N_{\rm D}^2 \rangle$ 
and 
$\langle N_{\rm H} N_{\rm D} \rangle$, 
as well as $m+1$ moments of any higher order $m>2$. 
The average rates of production of  
the H$_2$, HD and D$_2$ 
molecules on the surface of a single grain 
can be obtained by a proper summation over the third 
set of terms in the master equation 
(\ref{eq:NmicroHD}).  
These production rates 
can be expressed in terms of the first and second 
moments of the population size distribution 
\citep{Biham2001, Stantcheva2001,Stantcheva2002}: 
 
\begin{eqnarray} 
R_{\rm H_2} &=& A_{\rm H} \left(  \langle N_{\rm H}^2 \rangle  
- \langle N_{\rm H} \rangle \right) \nonumber \\ 
R_{\rm HD} &=& \left(A_{\rm H}+A_{\rm D}\right)  \langle N_{\rm H}N_ 
{\rm D} \rangle \nonumber \\ 
R_{\rm D_2} &=& A_{\rm D} \left(  \langle N_{\rm D}^2 \rangle  
- \langle N_{\rm D} \rangle \right).  
\label{eq:prodmoments} 
\end{eqnarray} 
 
\noindent 
The fact that the production rates depend on the second moments 
explains the failure of the rate equations in the limit in which 
the average surface populations of reactive species become $\le 1$. 
In the rate equations, which account only for the first moments,  
second moments are approximated by products of the corresponding 
first moments. The equations fail to account for the production 
rates when this approximation is no longer valid.  
 
The production rates 
given by Eqs. 
(\ref{eq:prodmoments}) 
can be obtained by direct integration of the master equation 
(\ref{eq:NmicroHD}) 
as a set of time dependent coupled differential equations, 
using a standard Runge-Kutta stepper. 
In this paper we focus on the  
production rates under steady state conditions. 
Therefore, the data collection is done only after  
the populations of the reactive species have reached 
their asymptotic sizes under the simulated conditions. 
In order to limit the number of equations, 
suitable cutoffs are imposed, either on the population  
size of each reactive specie or on the entire population 
of adsorbed atoms on the grain surface 
\citep{Stantcheva2001}. 
The production rates can also be obtained 
from Monte Carlo simulations, based on the master equation 
\citep{Charnley2001}. 
 
\section{The limit of very small grains}

In the limit of very small grains, one can reach the  
situation in which 
both  
$N_{\rm H} \ll 1$ 
and 
$N_{\rm D} \ll 1$. 
In this limit  
the formation rate of molecules is reduced and the 
dominant mechanism for removing atoms from the surface is 
thermal desorption.  
The formation of molecules can then be considered as a  
perturbation on top of the adsorption and desorption 
processes. 
This limit can be described by the master equation 
using cutoffs that do not allow more than a single atom 
on the grain. 
The master equation then 
consists of only three coupled equations, 
for  
$\dot P(0,0)$, 
$\dot P(1,0)$ 
and  
$\dot P(0,1)$. 
With these cutoffs, 
the average population sizes are related to the 
probabilities 
by 
 
\begin{eqnarray} 
\langle N_{\rm H} \rangle &=& P(1,0) \nonumber \\ 
\langle N_{\rm D} \rangle &=& P(0,1). 
\label{eq:N01} 
\end{eqnarray} 
 
\noindent 
Solving the three coupled equations for steady state 
conditions we obtain that 
 
\begin{eqnarray} 
\langle N_{\rm H} \rangle &=& F_{\rm H}/W_{\rm H} \nonumber \\ 
\langle N_{\rm D} \rangle &=& F_{\rm D}/W_{\rm D}. 
\label{eq:asymptN} 
\end{eqnarray} 
 
\noindent 
Consider the situation in which there is one atom on the surface 
of a grain. Since the grain is very small an atom can easily 
scan the whole surface before it desorbs, namely  
$A_{\rm I} \gg W_{\rm I}$. 
Thus, when a second atom 
is adsorbed, it is most likely to form a molecule with the atom 
that is already on the surface.  
Therefore, the production rate does not 
depend on the diffusion rates,  
a point noted by \citet{Tielens1982} and subsequently  
used to define the term ``accretion'' limit rigorously,  
as distinguished from the ``reaction'' limit. 
In the former limit, the production rates are given by 
the flux terms  
of H and D atoms 
in the master equation, 
multiplied by the probabilities 
of having a single atom (either H or D)  
on the surface, 
namely, 
$F_{\rm I} P(0,1)$  
and 
$F_{\rm I} P(1,0)$, 
where I $=$ H, D.  
Using Eqs. 
(\ref{eq:N01}), 
the production rates of molecules take the form 
 
\begin{eqnarray} 
R_{\rm H_2} &=& \langle N_{\rm H} \rangle F_{\rm H} \nonumber \\ 
R_{\rm HD} &=& \langle N_{\rm H} \rangle F_{\rm D}  
+ \langle N_{\rm D} \rangle F_{\rm H}  \nonumber \\ 
R_{\rm D_2} &=& \langle N_{\rm D} \rangle F_{\rm D}.  
\label{eq:prodsmallgr} 
\end{eqnarray} 
 
\noindent 
The HD molecule can form in two channels: an H atom on  
the surface reacts with a newly adsorbed D atom, 
or a D atom on the surface reacts with a newly adsorbed 
H atom.  
Due to the longer residence time of a D atom on the surface, 
the second channel is dominant. 
Therefore, the production ratio  
$R_{\rm HD} / R_{\rm H_2}$ 
is approximately equal to the ratio between the population sizes  
$\langle N_{\rm D}\rangle / \langle N_{\rm H} \rangle$ 
on the surface. 
This amounts to an enhancement of the HD production by 
a factor of $W_{\rm H}/W_{\rm D}$ compared to the flux 
ratio, 
namely 
 
\begin{equation} 
{R_{\rm HD} \over R_{\rm H_2}} =  
{W_{\rm H} \over W_{\rm D}} {F_{\rm D} \over F_{\rm H}}. 
\end{equation} 
 
\noindent 
The production rate of HD is thus enhanced with respect to that 
of H$_2$ by a factor given by the ratio of the desorption rates 
of H and D atoms. Since the desorption rates depend exponentially 
on the activation energy barriers for desorption 
$E_1(H)$ and $E_1(D)$, a small difference between these barriers 
can produce a large enhancement in HD production. 
The enhancement in D$_2$ production is given by 
 
\begin{equation} 
{R_{\rm D_2} \over R_{\rm H_2}} =  
{W_{\rm H} \over W_{\rm D}} \left({F_{\rm D} \over F_{\rm  
H}}\right)^2. 
\end{equation} 
 
\noindent 
giving rise to a similar enhancement compared to the expected  
production 
rate based on the ratio between the fluxes of D and H. 
 
\section{Simulations and results} 
 
To evaluate the production rates of  
H$_2$, HD and D$_2$ molecules 
we have  
used the master equation 
approach given 
by  
Eq.~(\ref{eq:NmicroHD}). 
The results presented in this paper were obtained by 
direct numerical integration of the set of differential 
equations until they reached steady state conditions. 
For a range of very small grains, namely $S<5000$, we also performed 
kinetic Monte Carlo simulations based on the master equation, and 
found perfect agreement between the two methods.  
 
In Fig. 1 we show 
the average population sizes of the atomic species  
on the grain surface, 
$\langle N_{\rm H} \rangle$ (solid line) 
and 
$\langle N_{\rm D} \rangle$ (dashed line), 
as a function of  
the number of adsorption sites, $S$,
and grain radius, $r$, 
under steady state conditions. 
The results are presented for a range of small grains on  
which the average number of H atoms is smaller than one 
while the average number of D atoms is of the order of 
$10^{-4}$ and less. 
Under these conditions 
the master equation is required in order to  
evaluate the production rates correctly, although the  
rate equations (dotted lines) are progressively better as the size of  
the grain increases.  Specifically, one can see that  
the rate equation approach appears to handle the average  
hydrogen abundance well for all grain sizes, while  
becoming much worse for the average deuterium abundance at  
small grain sizes.  One should note that the actual  
production rate of molecules does not depend on  
$\langle N_{\rm H} \rangle$ 
and $\langle N_{\rm D} \rangle$ 
only  
[see Eq. (\ref{eq:prodmoments})], so one cannot argue  
about the suitability of the rate equation method for  
the determination of production rates based on the results  
shown in  
Fig. 1. 
 
The production rates of 
H$_2$,  
HD 
and  
D$_2$,  
molecules 
per grain  
vs. $S$ and $r$ 
are shown in  
Fig. 2. 
Clearly, the production rates decrease as $S$ is reduced. 
For large grains that support sizable populations of H and D, 
and can be described by rate equations, 
the production rates per grain  
are proportional to the surface area, namely 
linear in $S$.  
In the limit of diminishing grain sizes, the production rates 
become quadratic in $S$,   
as can be seen by substitution of  
$\langle N_{\rm H} \rangle$ 
and 
$\langle N_{\rm D} \rangle$ into  
Eq. (\ref{eq:prodsmallgr}),  
which leads to a quadratic dependence on flux. 
The range of grain sizes presented in the simulation results 
reaches both the limit of large grains and the limit of small grains, 
covering the intermediate range between these two limits. 
 
In order to quantify the relative enhancement in the formation rate of 
deuterated molecules vs. H$_2$ we present in  
Fig. 3  
the ratios 
$R_{\rm HD}/R_{\rm H_2}$ 
and 
$R_{\rm D_2}/R_{\rm H_2}$ 
vs. $S$ and $r$. 
It is observed that as $S$ 
decreases these production ratios 
increase by over a factor of 20. 
In the limit of large grains,  
the production ratio converges towards 
the results of the rate equations (dotted line on right-hand side). 
In the limit of very small grains it converges towards the results 
of the asymptotic analysis presented in Sec. 5 (dotted line on left-hand side). 
 
To examine the dependence of the efficiency of the molecular formation 
on the grain size we show in  
Fig. 4 
the production rates per site  
of H$_2$, HD and D$_2$ 
on the grain as a function of $S$ and $r$. 
In the limit of large grains this quantity saturates and does not 
depend on the grain size.  
However, as the grain size decreases, the production rate per site 
goes down.  
The production rate of H$_2$ is reduced more quickly than 
that of the deuterated species. 
This gives rise to an enhancement in the relative 
production rate of HD and D$_2$ vs. H$_2$. 
 
To account for the production rate of molecular hydrogen in  
interstellar 
clouds one should integrate over the entire distribution of grain  
sizes 
in the cloud 
\citep{Mathis1977,Mathis1990,Mathis1996,Draine1984,Draine1985}. 
Consider a cloud in which the grain size distribution  
is given by  
$n_{\rm gr}(r)$ (cm$^{-4}$) 
in the range  
$r_{\rm min} < r < r_{\rm max}$,   
where  
$r_{\rm min}$   
and 
$r_{\rm max}$ 
are the lower and upper cutoffs of the distribution 
[here $n_{\rm gr}(r) dr$ (cm$^{-3}$) is the number density 
of grains of sizes in the range (r,r+dr)]. 
The formation rates   
${\cal{R}}_{\rm H_2}$,  
${\cal{R}}_{\rm HD}$  
and 
${\cal{R}}_{\rm D_2}$,  
(cm$^{-3}$ s$^{-1}$ ) 
in the cloud are given by 
 
\begin{equation} 
{\cal{R}}_{\rm H_2} = \int_{ r_{\rm min} }^{ r_{\rm max} }  
n_{\rm gr}(r) R_{\rm H_2}(r) dr, 
\label{eq:integ.prod.rateHH} 
\end{equation}

\begin{equation} 
{\cal{R}}_{\rm HD} = \int_{ r_{\rm min} }^{ r_{\rm max} }  
n_{\rm gr}(r) R_{\rm HD}(r) dr, 
\label{eq:integ.prod.rateHD} 
\end{equation} 
 
\noindent 
and 
 
\begin{equation} 
{\cal{R}}_{\rm D_2} = \int_{ r_{\rm min} }^{ r_{\rm max} }  
n_{\rm gr}(r) R_{\rm D_2}(r) dr, 
\label{eq:integ.prod.rateDD} 
\end{equation} 
 
\noindent 
respectively, 
where  
$R_{\rm H_2}(r)$, 
$R_{\rm HD}(r)$ 
and  
$R_{\rm D_2}(r)$ 
are the production rates  
per grain 
for grains of radius $r$. 
 
Observations support the assumption that the grain size distribution 
takes the power-law form 
\citep{Mathis1977,Weingartner2001} 
 
\begin{equation} 
n_{\rm gr}(r) = {K \over r^{\alpha}} 
\label{eq:power} 
\end{equation} 
 
\noindent 
in a range of sizes 
$r_{\rm min} < r < r_{\rm max}$, 
where 
 
\begin{equation} 
K = \frac{(\alpha-1)n_{\rm gr(total)}}{r_{\rm min}^{1-\alpha}-r_{\rm  
max}^{1-\alpha}}, 
\label{eq:K} 
\end{equation} 
 
\noindent 
and 
$n_{\rm gr(total)}$ (cm$^{-3}$) 
is the total density of grains, 
integrated over the whole range of grain sizes. 
To exemplify the role of the grain size distribution we  
analyze the case in which $\alpha=3$. 
This distribution has the special property that the mass of  
grains in the cloud is equally distributed within the range 
of grain sizes. 
The total surface area of all grains is distributed such that 
grains in the size range between $(r,r+dr)$ have a total surface 
area proportional to $1/r$. 
As a result, most of the surface area belongs to grains in the 
lower end of the size distribution. 
In the analysis we chose  
$r_{\rm min} = 4 \times 10^{-7}$ cm 
and 
$r_{\rm max} = 2 \times 10^{-5}$ cm.  
The parameter $K$ is chosen such that the density of  
large grains, of radii in the range 
$ 5 \times 10^{-6} < r < 2 \times 10^{-5}$ cm,  
is 
$n_{\rm gr(classical)} = 10^{-12} n_{\rm H}$.   
Integrating the grain size distribution over the 
entire range we obtain 
$n_{\rm gr(total)} = 1.66 \times 10^{-10} n_{\rm H}$.   
The minimum grain size is  
chosen to be larger than PAH molecules so that the problem  
of stochastic heating and desorption by photons does  
not dominate normal thermal evaporation. According to  
the calculations of \citet{Draine2001} and \citet{Li2001},  
however, grains of our minimum size may experience  
stochastic heating of temperatures up to 40 K at  
time intervals ($ \approx 10^{4}$ s$^{-1}$) somewhat  
shorter than the D evaporation time at the grain  
temperature chosen. Such heating would effectively  
shorten the grain lifetime of D atoms and reduce  
the isotope effect. The time scale between heating  
events could be lengthened by extinction in the cloud.  
 
To examine the contribution of grains 
of different sizes to the total production 
rates of H$_2$, HD and D$_2$ we show in  
Fig. 5  
the differential production rates  
given by the integrands of Eqs. 
(\ref{eq:integ.prod.rateHH}) 
and 
(\ref{eq:integ.prod.rateHD}) 
as a function of the grain radius $r$. 
Two competing effects play a role in this process. 
The production efficiency of molecules on the grain decreases as the  
grain 
size is decreased. However, small grains account for most of the  
surface 
area and thus provide a large fraction of the molecules that are  
produced. 
The relative enhancement in the production of deuterated species arises because  
their production efficiency decreases more slowly  
than that 
of H$_2$ as the grain size is reduced. 
 
Integrating over the range of grain sizes 
we obtain that the production rates are 
${\cal{R}}_{\rm H_2} = 3.9 \times 10^{-15}$ cm$^{-3}$ s$^{-1}$, 
${\cal{R}}_{\rm HD} = 1.2 \times 10^{-18}$ cm$^{-3}$ s$^{-1}$ 
and 
${\cal{R}}_{\rm D_2} = 1.5 \times 10^{-23}$ cm$^{-3}$ s$^{-1}$. 
Therefore, the ratio 
${\cal{R}}_{\rm HD}/{\cal{R}}_{\rm H_2} = 3.1 \times 10^{-4}$  
(${\cal{R}}_{\rm D_2}/{\cal{R}}_{\rm H_2} = 3.9 \times 10^{-9}$)  
is larger by about a factor of  
$44.1$  
($78.3$) 
than the expected 
value based on the ratio of the incoming fluxes, 
namely  
$F_{\rm D}/F_{\rm H}=10^{-5}/\sqrt{2}$. 
Observations indicate that the exponent  
in Eq.  
(\ref{eq:power}) 
is  
$\alpha = 3.5$ 
\citep{Weingartner2001}. 
Therefore, the contribution of very small grains  
may be even more dominant,  
leading to a larger enhancement in the formation of HD and D$_2$ 
vs. H$_2$ than in the analysis shown here.  
 
We will now compare our results with the standard rate law used 
for the production of molecular hydrogen in diffuse regions. 
This rate law assumes that  
all the grains are of radius 
$r = 10^{-5}$ cm 
and that their density is equal to 
$n_{\rm gr(classical)}$. 
It is also assumed that the efficiency is $\eta=1$, 
namely, all the atoms that collide with grains end up 
in molecules.  
Under these assumptions the production rate   
of H$_2$ molecules is given by 
${\cal{R}}_{\rm H_2} =  
n_{\rm H} v_{\rm H} n_{\rm gr(classical)} \sigma/2$  
cm$^{-3}$ s$^{-1}$ 
\citep{Hollenbach1971b}, 
or more conveniently, by 
 
\begin{equation} 
{\cal{R}}_{\rm H_2} = R({\rm H_2}) \  n_{\rm H} \ n, 
\end{equation} 
 
\noindent 
where 
$n_{\rm H}$ is the density of H atoms and $n$ is the total density 
of H nuclei in all forms 
(although we assume here that 
all the hydrogen in the gas is in the atomic form, namely 
$n=n_{\rm H}$).  
The parameter 
$R({\rm H_2})$ (cm$^3$ s$^{-1}$) 
is the effective rate coefficient given by   
 
\begin{equation} 
R({\rm H_2}) = 5 \times 10^{-17} \sqrt{T_{\rm gas}/300}. 
\label{eq:Rstandard} 
\end{equation} 
 
\noindent 
For the conditions considered here, $R({\rm H_2})=2.74 \times 10^{-17}$  
cm$^3$ s$^{-1}$ 
and 
${\cal{R}}_{\rm H_2} = 6.85 \times 10^{-14}$ 
cm$^{-3}$ s$^{-1}$. 
This result is 
about an order of magnitude higher than 
our result, which is consistent with 
$R({\rm H_2})=1.56 \times 10^{-18}$  
cm$^3$ s$^{-1}$. 
The reason for this apparent discrepancy is the low 
production efficiency of H$_2$ molecules  
under these conditions, 
which on large grains 
is 
$\eta=0.09$ 
(given by the rate equations)  
and is even lower for small grains.   
The production rate of HD molecules 
(cm$^{-3}$ s$^{-1}$) 
can be expressed in a
similar way, by
 
\begin{equation} 
{\cal{R}}_{\rm HD} = R({\rm HD}) \  n_{\rm D} \ n, 
\end{equation} 
 
\noindent
where
$R({\rm HD})$ 
(cm$^3$ s$^{-1}$) 
is the effective rate coefficient for this reaction.   
The values of the rate coefficients
$R({\rm H_2})$ 
and
$R({\rm HD})$ 
under the conditions studied here,
are shown in Table 1,
for 
$\alpha=3$ and $3.5$
and for several grain temperatures
between 12 and 20 K. 

In Fig. 6 we present  
the temperature dependence of the production rates 
of H$_2$ and HD molecules. 
The production rates  
${\cal{R}}_{\rm H_2}$, 
obtained from the rate equations, 
are shown by the solid lines 
for the case that the exponent of the grain 
size distribution is $\alpha = 3.5$ (upper solid line) 
and for $\alpha=3$ (lower solid line). 
The master equation results for these two cases 
are also shown (by $\circ$ and $\times$, respectively). 
The production rates  
${\cal{R}}_{\rm HD}$, 
obtained from the rate equations, 
are shown by the dashed lines, for 
$\alpha = 3.5$ (upper dashed line) 
and for $\alpha=3$ (lower dashed line). 
The master equation results  
are also shown (by $\circ$ and $\times$, respectively). 
These graphs provide much insight on the enhancement of 
HD production. 
They show that for grain temperatures lower than 
about 16 K, the production efficiency is  
$\eta \simeq 1$ 
and the rate coefficient $R({\rm H_2})$ is consistent with 
Eq.  
(\ref{eq:Rstandard}). 
At this temperature range there is no enhancement in 
the production of HD molecules. 
The enhancement appears in a range of temperatures, 
roughly between 16 K and 20 K, where the H$_2$  
production rate quickly decreases  
as $T$ increases 
while the HD  
production rate remains steady. 
Above 20 K the production rates of both H$_2$ and 
HD are diminished. 
 
In Fig. 7 we present the ratio,  
${\cal{R}}_{\rm HD}/{\cal{R}}_{\rm H_2}$, 
of the production rates 
of HD and H$_2$ molecules as a function of the grain  
temperature, 
as obtained from the rate equations (solid line), 
and the master equation for  
$\alpha=3$ ($\times$) 
and 
$\alpha=3.5$ ($\circ$). 
The enhancement ratio increases sharply as the temperature 
is raised above 16 K and reaches a peak around 20 K. 
A significant enhancement is observed on large grains,  
for which the rate equation results apply. 
When the grain size distribution is taken into account, 
further enhancement is obtained due to the contribution 
of the small grains, on which the enhancement ratio is  
even larger. 
 
The dependence of the ratio of the production rates 
$R_{\rm HD}/R_{\rm H_2}$ 
on the grain size $S$ and radius $r$ 
is shown in Fig. 8 
for several grain temperatures. 
At 14 K there is no significant enhancement 
for any grain size. 
At 16 and 18 K the enhancement strongly depends on 
the grain size and increases sharply as the grain 
size decreases. 
At 20 K there is a large enhancement for grains of 
all sizes. However, at this temperature the total production 
of both H$_2$ and HD is very small, 
thus making little contribution to the gas phase abundances.

\section{Discussion} 
 
In this paper we have used a simple description of the grains 
as spherical objects. In practice the grains are expected to 
be amorphous with rough surfaces and may include pores.  
These features may increase the surface area of the grain 
and introduce a distribution of diffusion and desorption 
barriers. However, no solid data are available to quantify 
such effects. Moreover, the analysis of experimental results on  
molecular hydrogen production on olivine and amorphous 
carbon surfaces 
\citep{Pirronello1997a,Pirronello1997b,Pirronello1999} 
did not require more than one energy barrier 
for each process 
\citep{Katz1999}. 
 
The analysis presented in this paper is restricted to the case 
in which the mobility of adsorbed atoms on the surface is due 
to thermal hopping. Tunneling of H and D atoms on the surface 
is not included since experimental results  
indicate that tunneling  
does not play a significant role 
in the formation of molecular 
hydrogen under astrophysically relevant conditions  
\citep{Pirronello1997a,Pirronello1997b,Pirronello1999}. 
In a recent analysis of the results of these 
experiments, chemisorption processes were taken into account 
\citep{Cazaux02}. 
The study of chemisorption processes is  
motivated by observations of molecular hydrogen in photon-dominated 
regions in which the typical grain temperature is around 40 K, 
making the production of H$_2$ molecules by physisorbed H 
atoms hopelessly inefficient. 
 
The mechanism presented here for an enhanced production of HD  
and D$_2$ molecules, vs. H$_2$, takes place in a window of 
grain temperatures.  
For the surface parameters used here this temperature window is 
between 16 - 20 K. In this temperature range the production efficiency  
of H$_2$ is reduced, while the efficiency of HD and D$_2$ production  
remains steady.  
The effect is observed for grains of all sizes, but the relative 
enhancement is larger on small grains.  It will be interesting to see 
if the enhancement plays a significant role in chemical models of diffuse 
and translucent clouds; relevant model calculations are planned by Le Petit 
and Roueff (private communication).

\section{Summary}  
 
The formation of  
H$_2$, HD and D$_2$ molecules 
on dust grain surfaces was studied 
using the master equation approach. 
The production of these molecules 
was assumed to occur via the Langmuir-Hinshelwood (diffusive) mechanism. 
It was shown that for surface temperatures 
in excess of 15 K,  
the fractions of HD and D$_2$ molecules in the population of 
hydrogen molecules that are produced are significantly enhanced  
compared with their expected ratio based on the gas phase abundances. 
The enhancement takes place on grains of all sizes, but is largest 
on the smallest grains considered until the temperature reaches 
20 K. 
It arises because of the assumption 
that D atoms are bound slightly more strongly to grains than are H atoms.  
At a surface temperature of 18 K, with binding energies and rates of 
diffusion based on laboratory experiments, 
 the enhancement is found to be a factor of 44.1 (78.3) for the 
production of HD (D$_{2}$)with the assumption that the grain size distribution 
goes as $r^{-3}$.  Although the enhancement is even larger at higher temperatures, 
the absolute rates of production for H$_{2}$ and its isotopomers become too small 
to be of importance.  Use of the master equation method can lead to somewhat  
greater enhancements (factor of a few) than the simple rate equation method.  
 
While gas-phase mechanisms  
\citep{Watson1973} 
may account for the observed abundance 
of HD, it is possible that surface mechanisms such as the one 
presented here also play a role in the production of deuterated molecules. 
Recent observations have uncovered a variety of 
multiply deuterated molecular species in the 
interstellar medium 
such as D$_2$CO and ND$_3$ 
\citep{Ceccarelli2002,vanderTak2002}.  
Gas phase reaction networks that lead to the formation of such  
multiply 
deuterated species have been studied and their efficiency  examined 
\citep{Charnley1997,Rodgers2001, Roberts2002, Roberts2003}. 
The picture in pre-stellar cores is that gas-phase  
fractionation can explain much of what is observed,  
but the picture in protostellar cores, where grain  
mantles tend to be evaporated back into the gas, is  
that surface reactions play an important role in the  
fractionation leading to multi-deuterated species \citep{Stantcheva2002}. 
It will be interesting to see if the surface mechanism presented 
here may be extended to  
produce multiply deuterated species.   
Preliminary work on this matter has  
been undertaken by \citet{Caselli2002} and \citet{Stantcheva2003}. 
 
\section{acknowledgments} 
This work was supported by the Adler Foundation for Space Research of  
the  
Israel Science Foundation.  E. H. wishes to thank the National  
Science Foundation (U. S.) for support of his research programme in  
astrochemistry.

\newpage 
\clearpage 

\begin{table} 
\caption{The rate coefficients $R({\rm H_2})$ and
$R({\rm HD})$ (cm$^{3}$ s$^{-1}$) for several 
grain temperatures between 12 and 20 K
and for grain size distributions characterized by
$\alpha=3$ and $3.5$. 
} 
\begin{tabular}{cllll} 
T & 
\multicolumn{2}{c}{$R({\rm H_2})$} & 
\multicolumn{2}{c}{$R({\rm HD})$}\\ 
\hline 
&\multicolumn{1}{c}{$\alpha=3$}&\multicolumn{1}{c}{$\alpha=3.5$}& 
\multicolumn{1}{c}{$\alpha=3$}&\multicolumn{1}{c}{$\alpha=3.5$}\\ 
\cline{2-5} 
12 & $4.31 (-17)$ & $8.35 (-17)$ & $6.1 (-17)$ & $1.18 (-16)$\\ 
14 & $4.40 (-17)$ & $8.18 (-17)$ & $6.36 (-17)$ & $1.2 (-16)$\\ 
16 & $2.29 (-17)$ & $3.03 (-17)$ & $6.33 (-17)$ & $1.19 (-16)$\\ 
18 & $1.56 (-18)$ & $1.67 (-18)$ & $4.85 (-17)$ & $7.53 (-17)$\\ 
19 & $2.22 (-19)$ & $2.46 (-19)$ & $2.82 (-17)$ & $3.59 (-17)$\\ 
20 & $3.11 (-20)$ & $3.92 (-20)$ & $7.19 (-18)$ & $9.19 (-18)$\\ 
\hline 
\end{tabular} 
\end{table} 

\newpage
\clearpage 
 
\begin{figure} 
\vspace{12pt} 
\epsfxsize=14cm 
\epsffile{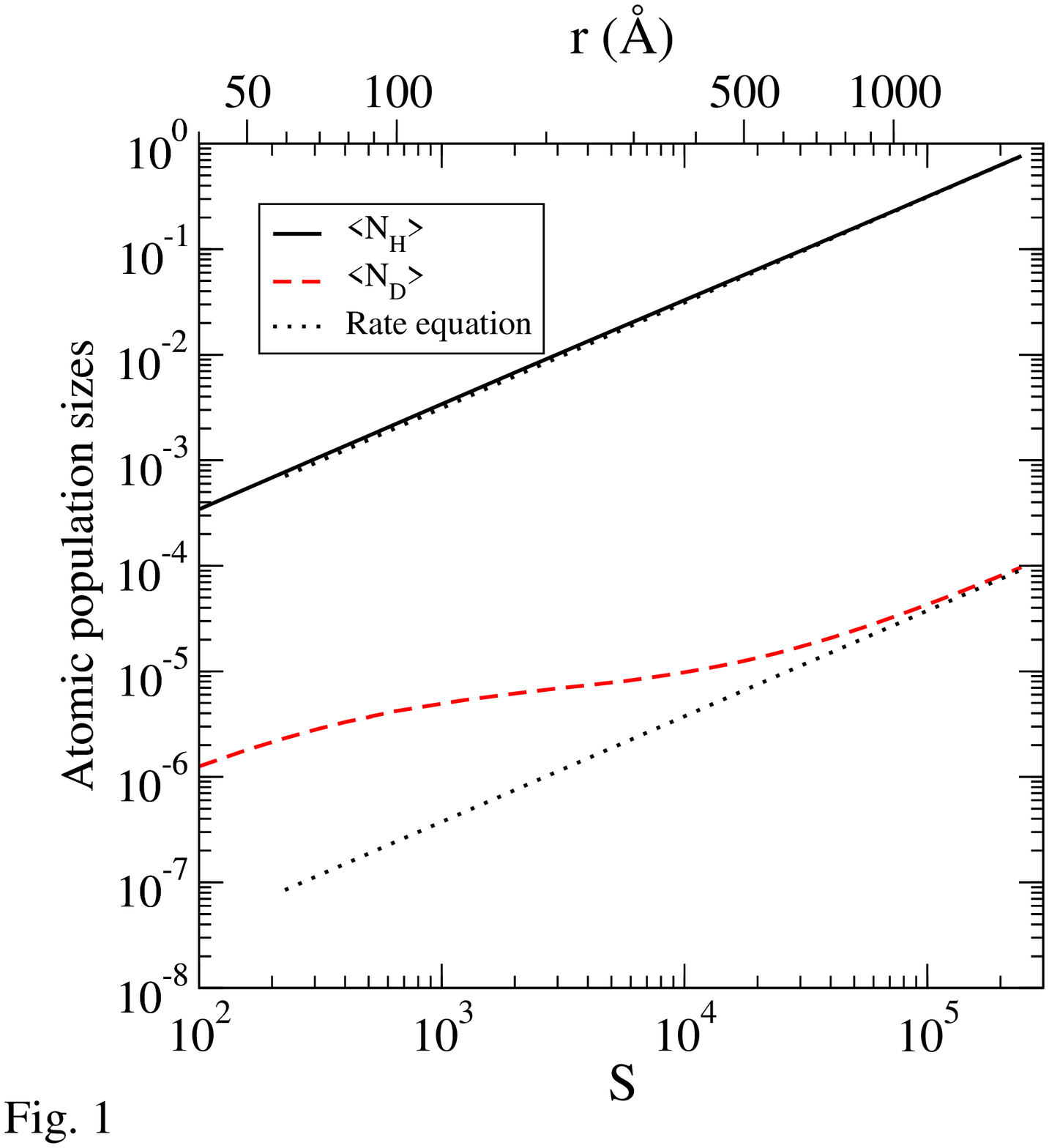} 
\caption{The average number of H atoms,  
$\langle N_{\rm H} \rangle$ (solid line), 
and the average number of D atoms, 
$\langle N_{\rm D} \rangle$ (dashed line) 
on the surface of a grain that has $S$ adsorption  
sites, as a function of $S$ and grain radius $r$, 
obtained from direct integration of the master equation. 
The corresponding rate equation results are shown by dotted lines. 
} 
\end{figure} 
 
\begin{figure} 
\epsfxsize=14cm 
\epsffile{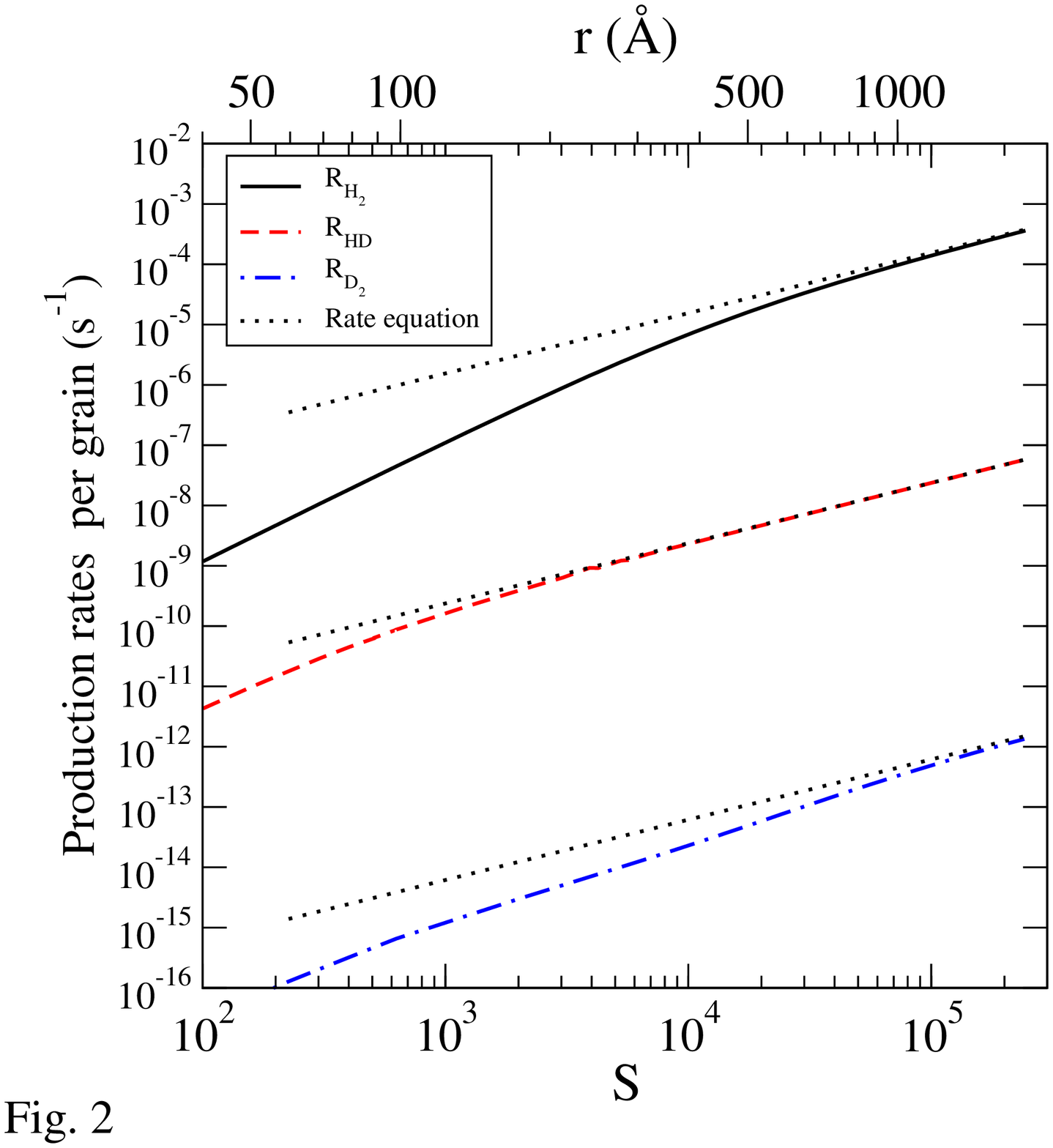} 
\caption{ 
The production rates per grain of H$_2$ (solid line), HD (dashed line)  
and D$_2$ (dashed-dotted line) 
molecules as a function of the number of adsorption sites, $S$, on  
the  
grain surface and grain radius $r$.  
In the limit of large grains 
the production rates approach a linear dependence on $S$, while in 
the limit of very small grains they approach a  
quadratic dependence on $S$. 
} 
\end{figure} 
 
\begin{figure} 
\epsfxsize=14cm 
\epsffile{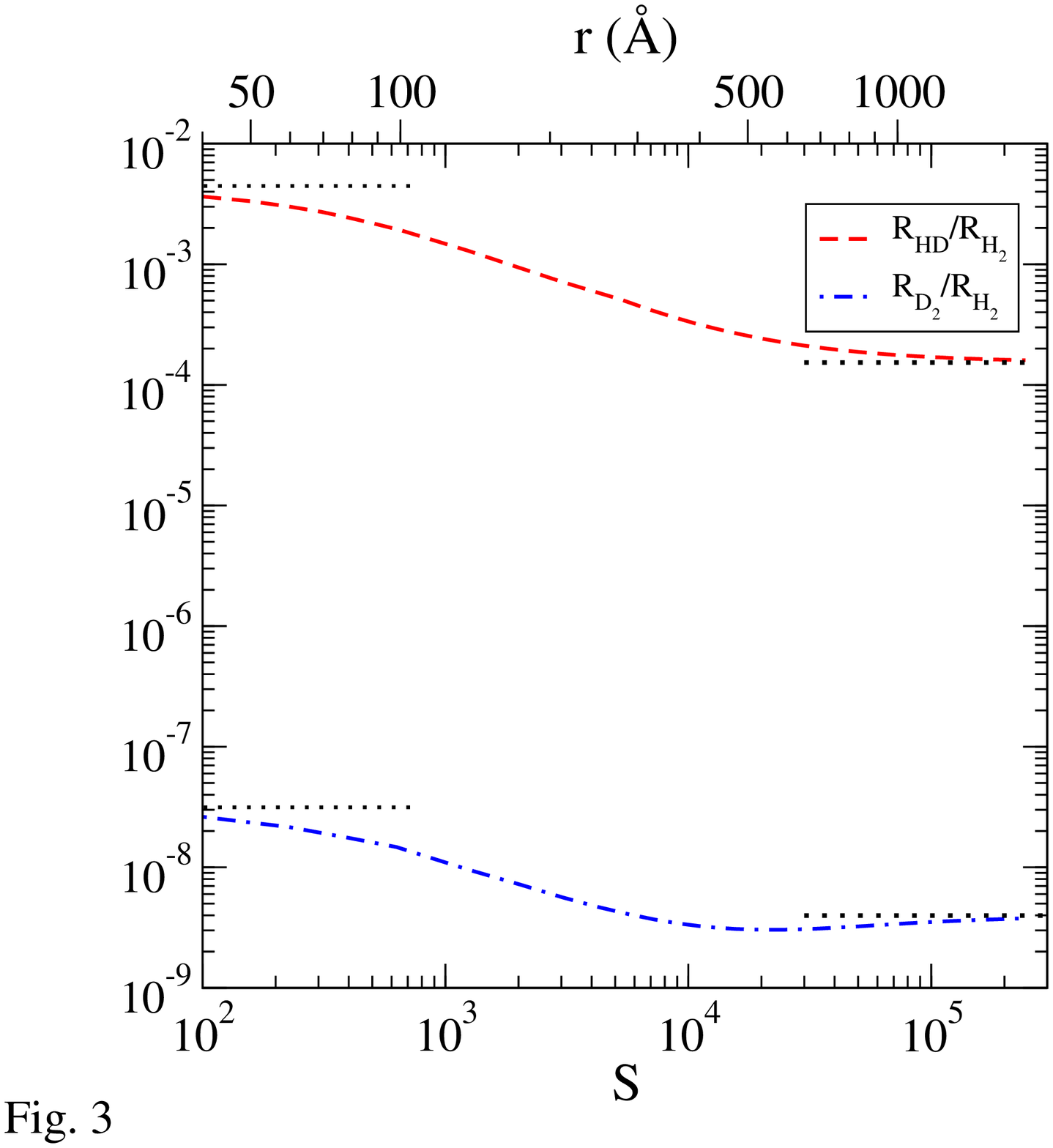} 
\caption{The ratio between the production rates of HD [D$_2$] and H$_2$  
molecules, 
$R_{HD}/R_{H_2}$ (dashed line) 
[$R_{D_2}/R_{H_2}$ (dashed-dotted line)] 
vs. the number of adsorption sites $S$ on the grain surface and grain 
radius $r$.  
In the limit of large grains both ratios converge to the rate equation  
results (dotted lines on the right-hand side).  
As the grain size goes down, these ratios  
increase 
by over a factor of 20 until they converge to the results of the 
asymptotic analysis for the limit of very small grains  
(dotted lines on the left-hand side). 
} 
\end{figure} 
 
\begin{figure} 
\epsfxsize=14cm 
\epsffile{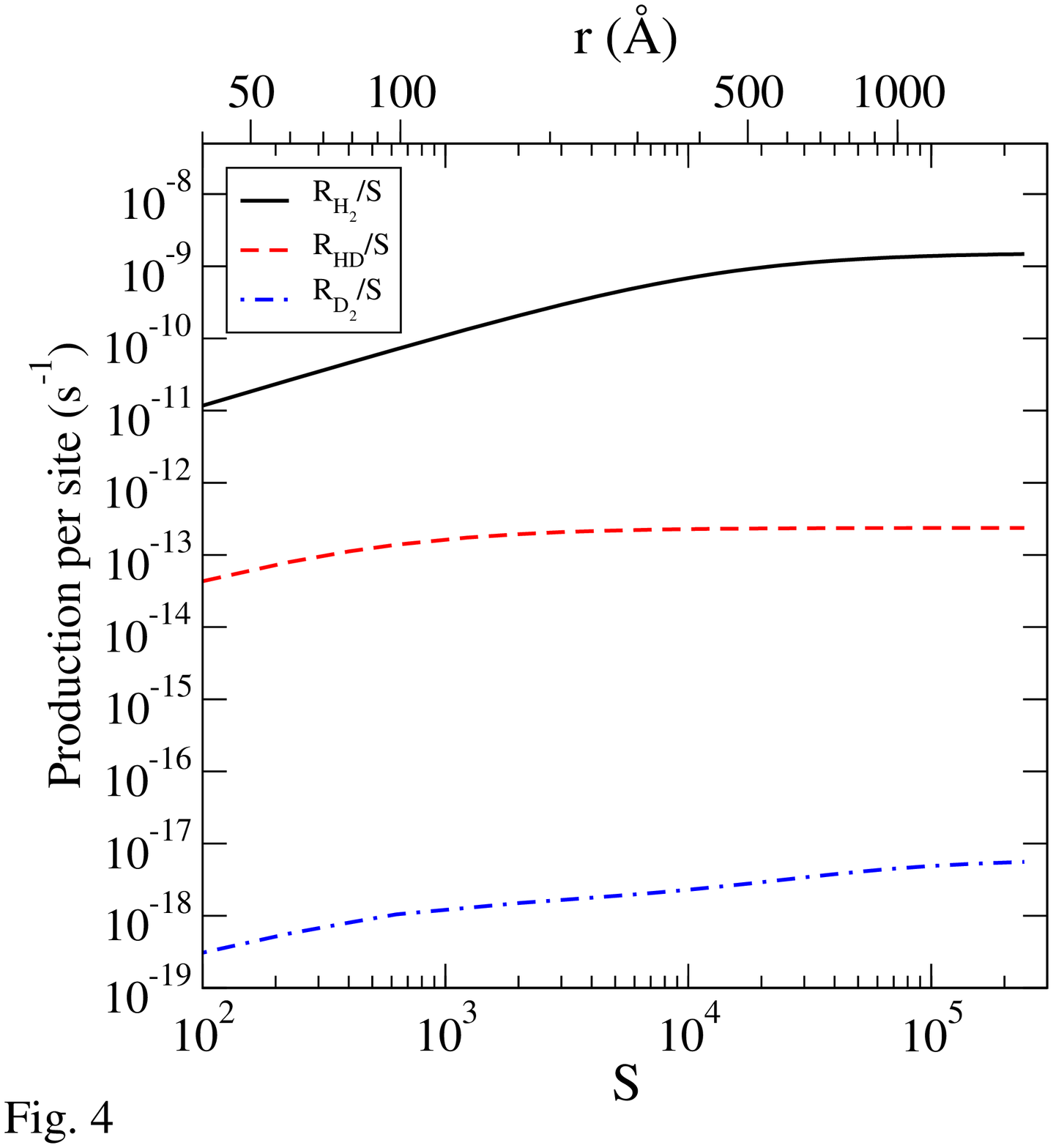} 
\caption{The production rates  
of 
H$_2$,  
HD  
and 
D$_2$,  
molecules  
per site on the surface of a grain that has $S$ 
adsorption sites, 
as a function of $S$ and grain radius $r$. 
These curves quantify the dependence of the efficiency  
of molecular production per surface area on the 
size of the grain. 
While the production rate of H$_2$ declines  
sharply as the grain size decreases, there are 
more moderate reductions in the production rates 
of HD and $D_2$. 
} 
\end{figure} 
 
\begin{figure} 
\epsfxsize=14cm 
\epsffile{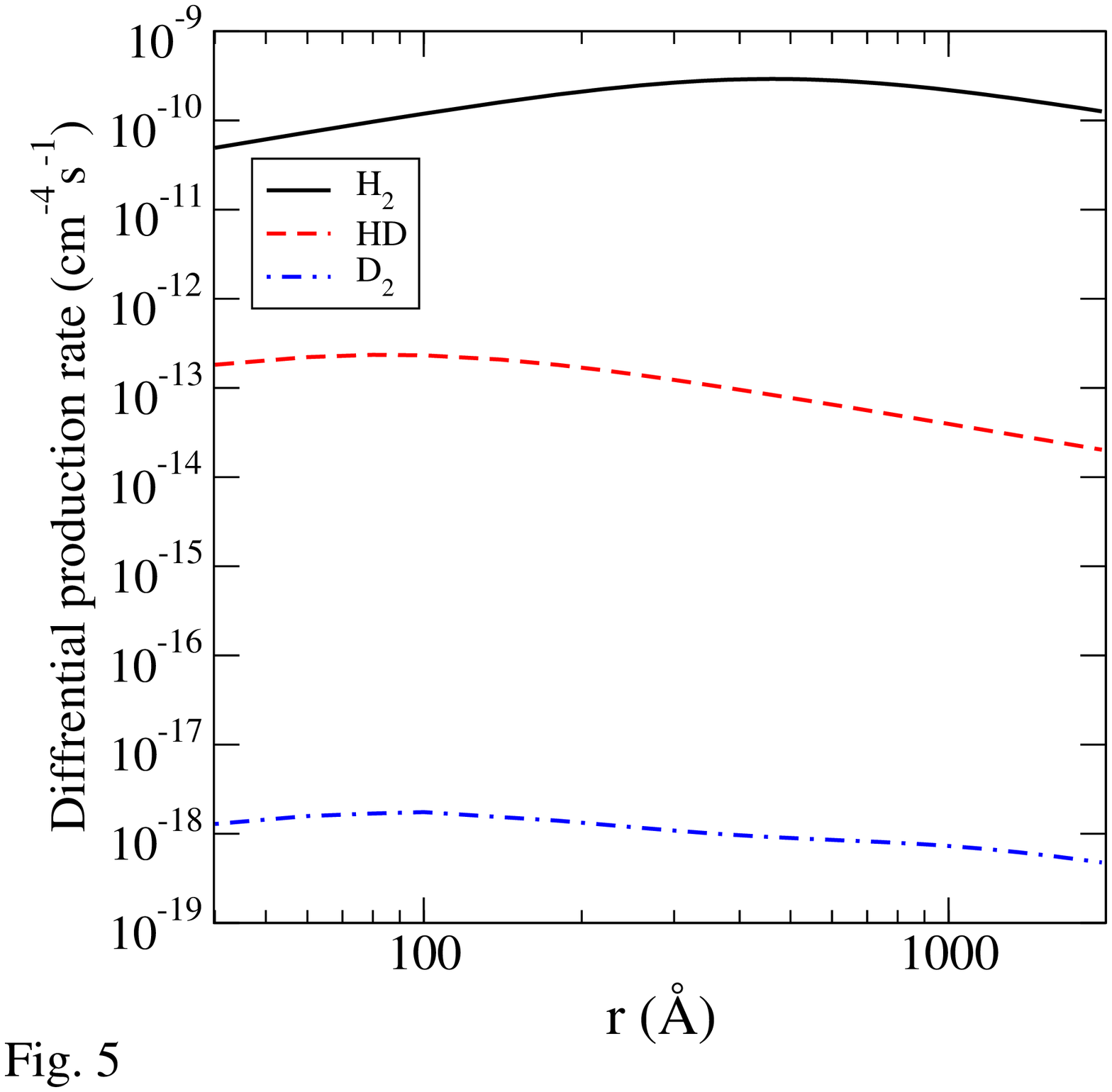} 
\caption{ 
The differential production rates of 
H$_2$,  
HD  
and 
D$_2$  
molecules 
[given by the integrands in Eqs. 
(\ref{eq:integ.prod.rateHH}), 
(\ref{eq:integ.prod.rateHD}) 
and  
(\ref{eq:integ.prod.rateDD}), 
respectively] 
on the surfaces of grains of radii in the range 
$r,r+dr$.  
} 
\end{figure}

\begin{figure} 
\epsfxsize=14cm 
\epsffile{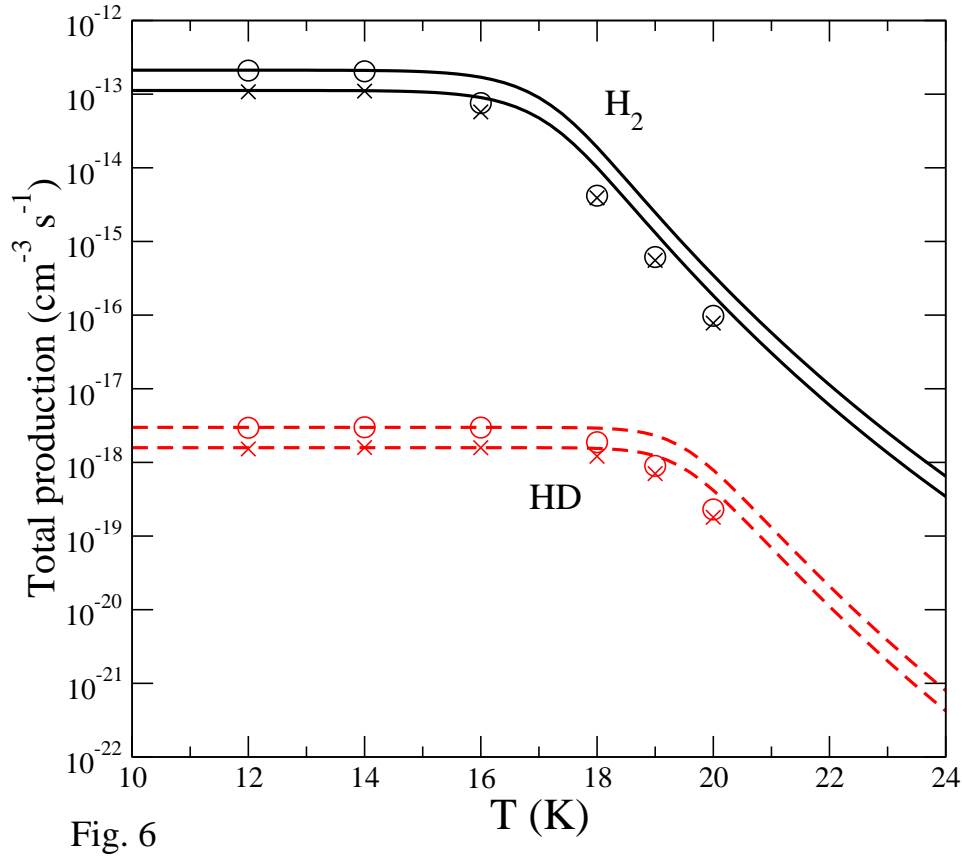} 
\caption{ 
The production rates, obtained from the rate equations, 
of H$_2$ (solid lines) and HD (dashed lines) molecules 
as a function of the grain temperature, for 
$\alpha = 3.5$ (upper line in each pair) 
and for $\alpha=3$ (lower line in each pair). 
The master equation results for these two values of $\alpha$ 
are also shown (by $\circ$ and $\times$, respectively). 
} 
\end{figure}

\begin{figure} 
\epsfxsize=14cm 
\epsffile{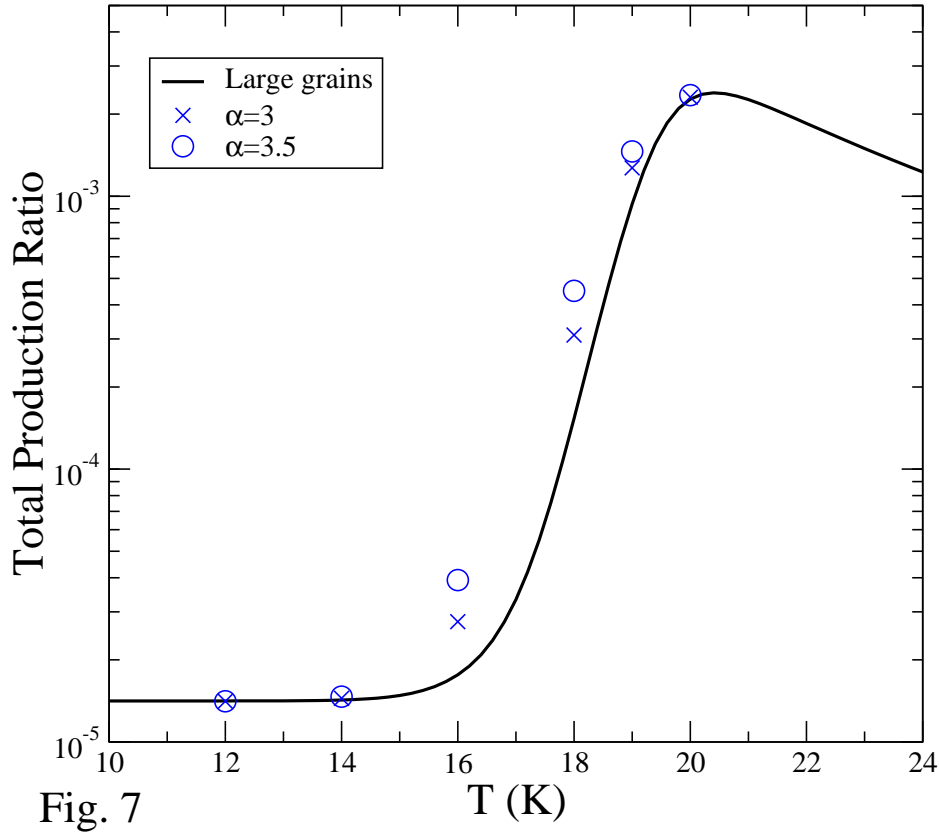} 
\caption{ 
The ratio between the production rates 
of HD and H$_2$ molecules as a function of the grain  
temperature, 
as obtained from the rate equations (solid line), 
and the master equation for  
$\alpha=3$ ($\times$). 
and 
$\alpha=3.5$ ($\circ$). 
The enhancement ratio increases sharply as the temperature 
is raised above 16 K and reaches a peak around 20 K. 
The enhancement exists for all grain sizes 
with further enhancement in the case of small grains. 
} 
\end{figure}

\begin{figure} 
\epsfxsize=14cm 
\epsffile{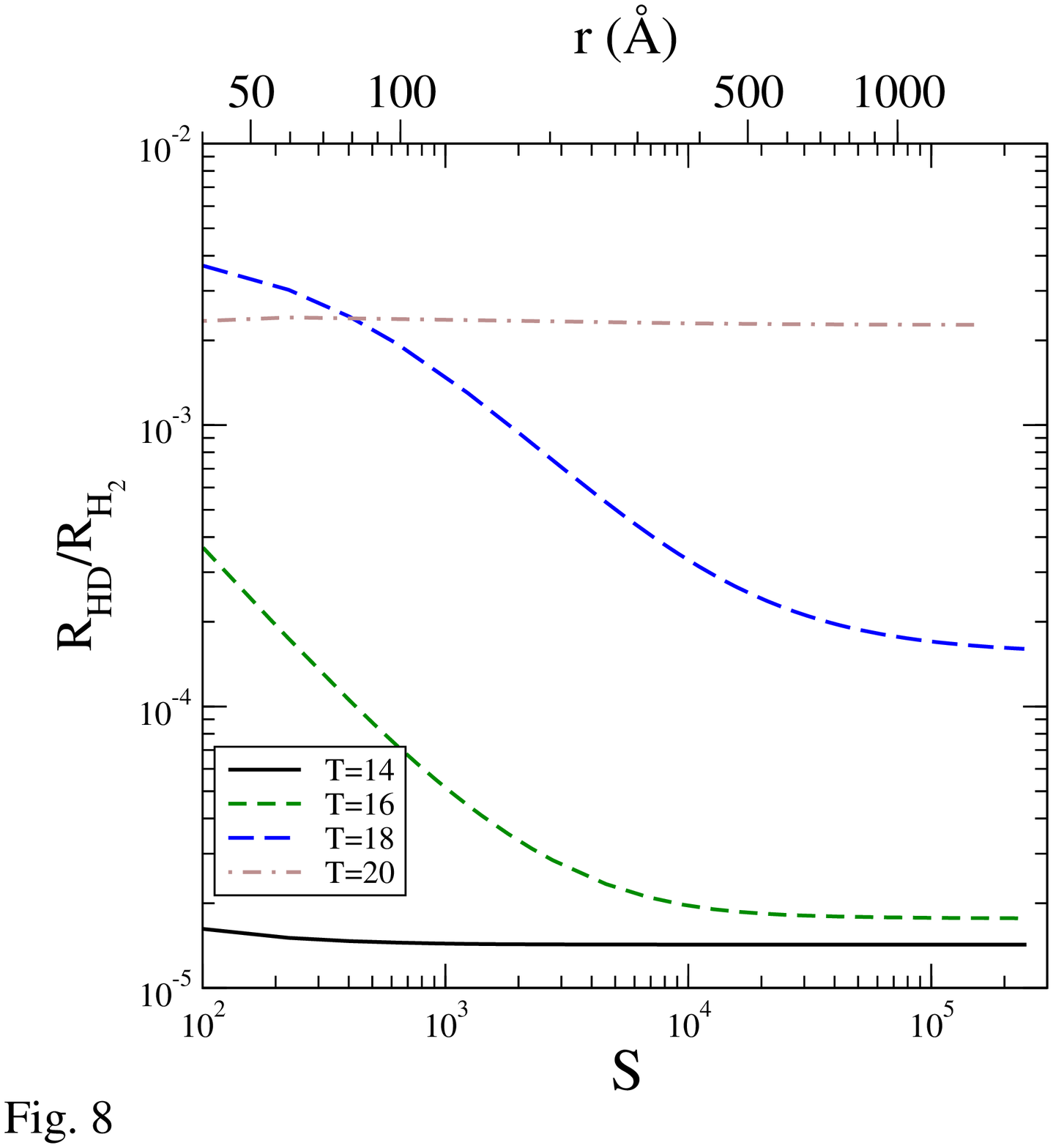} 
\caption{ 
The ratio of the production rates, 
$R_{\rm HD}/R_{\rm H_2}$, 
vs. grain size $S$ and radius $r$, 
for several grain temperatures. 
At $T=14$ K there is no significant enhancement. 
At $T=16$ and 18 K the enhancement strongly depends on 
the grain size and increases sharply as the grain 
size is reduced. 
At $T=20$ there is a large enhancement for grains of 
all sizes. However, at this temperature the total production 
of both H$_2$ and HD is very small. 
} 
\end{figure} 

\newpage
\clearpage

\appendix 
 
\section[]{Isotope effects in related experiments} 
 
Isotope effects resembling the type studied here were observed 
experimentally on various 
surfaces including metals  
\citep{Hoogers1995}  
and silicon 
\citep{Koehler1988}.  
However, these experiments were done at much higher temperatures than 
those relevant for molecular hydrogen formation. 
Also, the surfaces on which isotope effects were studied are more  
ordered than the 
astrophysically relevant surfaces. 
Therefore, it may not be possible to draw conclusions from these  
experiments 
for astrophysically relevant surface processes.  
 
A crucial feature in the recent experiments 
on molecular hydrogen formation is the use of two beams, 
of H and D atoms, and the measurement of the HD production rate 
\citep{Pirronello1997a,Pirronello1997b,Pirronello1999}. 
The use of HD molecules improves the  
signal-to-noise ratio in the experiment and allows reliable 
measurement of the production rate. 
In the analysis of the experimental results it was assumed, 
for simplicity, that there is no isotope effect, 
namely, that the diffusion and desorption barriers for H 
and D atoms are identical 
\citep{Katz1999}. 
This assumption was made in order to limit the number of 
free parameters used in the fitting.   
Under the experimental conditions this assumption had little 
effect on the results of the analysis. 
To explain this issue, let us recall the experimental procedure. 
The experiment followed the standard temperature programmed desorption 
(TPD) procedure, which consists of two stages. 
In the first stage, the surface was irradiated by two beams: 
one of H atoms and the other of D atoms. The beams were of 
(approximately) equal fluxes and lasted for an equal irradiation 
time, giving rise to equal coverages of H and D atoms. 
In the second stage the surface temperature was raised 
at a constant rate and the production rate of HD molecules was   
recorded using a mass spectrometer. 
 
The analysis of the experimental results was based on the fact 
that the surface was of macroscopic size,  
so that rate equations of the form  
(\ref{eq:NHgrain}) 
can account correctly for the atomic populations 
on the surface and for the production rates of molecules. 
In the first stage of the experiment the fluxes were 
$F_{\rm H} = F_{\rm D}$, 
while the diffusion and desorption rates were practically zero, 
due to the low surface temperature. 
In the second stage the two beams were turned off, 
while the diffusion and desorption coefficients 
quickly increased due to the heating of the surface. 
 
The isotope effect studied here gives rise to a  
small difference in the energy barriers for  
diffusion of H and D atoms, 
$E_0({\rm H}) < E_0({\rm D})$, 
and a larger difference 
in the desorption energies, 
where 
$E_1({\rm H}) < E_1({\rm D})$. 
We will now examine how these differences 
would affect the results of the TPD experiment. 
We first consider the effect of the difference in the diffusion barriers. 
The rate of formation of HD molecules is described by  
Eq. 
(\ref{eq:HD}), 
where the diffusion rates of the H and D atoms sum up.  
It is thus clear that even if the D atoms   
are immobile, this gives rise only to about a factor 
of 2 in the production rate compared to the case in which 
$A_{\rm D} = A_{\rm H}$. 
Such differences have little effect on the energy barrier 
for diffusion,  
$E_0$,  
obtained from fitting the experimental TPD curve. 
This is because 
$E_0$ 
appears in the exponent 
of Eq. 
(\ref{eq:Alpha}). 
Therefore, when the experimental TPD curves are fitted by a rate equation 
model that includes a single energy barrier for diffusion, 
$E_0$, the result is very close to the lower diffusion barrier 
among the H and D,  
namely to 
$E_0({\rm H})$. 
 
Consider the case in which the desorption barrier of D atoms, 
$E_1({\rm D})$,  
is larger than the desorption barrier 
of H atoms, 
$E_1({\rm H})$. 
In this case, as the temperature is raised during a TPD 
experiment, D atoms desorb more slowly than H atoms. 
The formation rate of HD molecules is thus limited by 
the availability of H atoms.  
When the resulting TPD curves are fitted by a rate 
equation model that includes only a single energy barrier 
for desorption, 
$E_1$, the fitted value turns out to be a very good 
approximation to  
$E_1({\rm H})$. 
 
It thus turns out that slower diffusion and desorption rates  
of D atoms vs. H atoms, have 
little effect on the TPD curves that describe the 
rate of formation of HD molecules. When these TPD curves 
are fitted by a rate equation model  
with a single diffusion barrier,  
$E_0$,  
and a single desorption barrier, 
$E_1$, 
these barriers 
provide good approximations to  
$E_0({\rm H})$ 
and  
$E_1({\rm H})$. 
This conclusion is not limited to the case in which the 
experiment is done on a macroscopic surface. It is expected 
to be valid even if the experiment is done on a powder that  
consists of small grains. 
 
The origin of the isotope effect on dust grains in the interstellar 
medium is the long residence time of D atoms on the grain  
compared with H atoms.  
As a result, an (H or D) atom that hits a grain is more likely to   
encounter a D atoms on the grain than expected based on the gas 
phase abundance ratio. 
In the TPD experiments all the H and D atoms are deposited on the surface 
at the initial stage of the experiment, when the surface temperature is  
low and the adsorbed atoms are immobile. 
During the heating stage, when the adsorbed atoms become mobile, 
the beams are turned off and 
there is no fresh supply of H or D atoms.  
Therefore, the D atoms that remain on the surface  
after most of the H atoms are desorbed 
tend to form D$_2$ molecules.  
The isotope effect is thus expected to affect  
the TPD curve of D$_2$ molecules. 
However, it does not affect the TPD 
curve of HD molecules, which is the one that was measured 
in the experiments.

\end{document}